# Exploring spatiotemporal network transitions in task functional MRI

Gregory Scott*, Peter J. Hellyer*, Adam Hampshire, Robert Leech

*The Computational, Cognitive and Clinical Imaging Laboratory, Division of Brain Sciences, Imperial College London, Hammersmith Hospital, Du Cane Road, W12 0NN*

*These authors contributed equally to the work presented in this report.

*Corresponding Author:* Robert Leech, Computational, Cognitive and Clinical Neuroimaging Laboratory, Hammersmith Hospital, London, W12 0NN, United Kingdom.

## Keywords

## Acknowledgement

Data were provided by the Human Connectome Project, WU-Minn Consortium (Principal Investigators: David Van Essen and Kamil Ugurbil; 1U54MH091657) funded by the 16 NIH Institutes and Centers that support the NIH Blueprint for Neuroscience Research; and by the McDonnell Center for Systems Neuroscience at Washington University.

# Abstract

A critical question for cognitive neuroscience regards how transitions between cognitive states emerge from the dynamic activity of functional brain networks. However, current methodologies cannot easily evaluate both the spatial and temporal changes in brain networks with cognitive state. Here we combine a simple data reorganization with spatial ICA, enabling a spatiotemporal ICA (stICA) which captures the consistent evolution of networks during onset and offset of a task. The technique was applied to FMRI datasets involving alternating between rest and task and to simple synthetic data. Starting and finishing time-points of periods of interest (anchors) were defined at task block onsets and offsets. For each subject, the ten volumes following each anchor were extracted and concatenated spatially, producing a single 3D sample. Samples for all anchors and subjects were concatenated along the fourth dimension. This 4D dataset was decomposed using ICA into spatiotemporal components. One component exhibited the transition with task onset from a default mode network (DMN) becoming less active to a fronto-parietal control network (FPCN) becoming more active. We observed other changes with relevance to understanding network dynamics, e.g., the DMN showed a changing spatial distribution, shifting to an anterior/superior pattern of deactivation during task from a posterior/inferior pattern during rest. By anchoring analyses to periods associated with the onsets and offsets of task, our approach reveals novel aspects of the dynamics of network activity accompanying these transitions. Importantly, these findings were observed without specifying *a priori* either the spatial networks or the task time courses.

# Introduction

A critical question for cognitive neuroscience regards how transitions between cognitive states emerge from the shifting balance in activity of functional brain networks. According to the dominant model, cognitively challenging tasks, such as a visually cued working memory task, evoke well-recognized patterns of activation and deactivation with functional neuroimaging (Fox et al., 2005; Smith et al., 2009). For example, frontal and parietal regions form the fronto-parietal control network (FPCN), which typically activates with sensory and motor systems during many cognitively challenging tasks (Dosenbach et al., 2007; Vincent et al., 2008). In contrast, activity within posterior cingulate, ventromedial frontal and lateral parietal cortex, forming the default mode network (DMN), is typically increased at rest, when thought is unfocused or focused internally (Buckner et al., 2008; Raichle et al., 2001), and is typically decreased during externally focused tasks (Chang and Glover, 2010; Fox et al., 2005). The perceived dichotomy of activity between the DMN and FPCN encourages a caricature of network dynamics where an active DMN represents a basal state of brain activity that is transiently interrupted or "turned off" during task conditions, whilst the FPCN (alongside sensory and motor systems) activates. When the active, task state finishes, the FPCN and sensori-motor systems turn off, leaving the brain to return to its basal state of DMN-predominant activity.

However, this dichotomous model may be overly-simplistic. There is evidence that the dynamics of the interactions between functional brain networks supporting cognitive functions are much more complex. For example, functional connectivity between brain networks, even at rest, has been shown to be non-stationary, fluctuating over time (Chang and Glover, 2010). Further, there is increasing evidence that cognitive control relies on transient competitive or cooperative interactions between brain networks (Shmuel et al., 2002). Therefore, in the context of a given cognitive task, there may be multiple different network transitions, and these transitions could involve more complexity than an on-off cycle between the DMN and the FPCN. Further, it has been proposed that the spatial composition of the networks themselves may be non-stationary and may adapt to cognitive demands (Leech et al, 2014). For example, the spatial pattern of relative DMN deactivation evoked by even quite similar cognitive tasks can vary substantially (Harrison et al., 2011; Leech et al., 2011; Leech et al., 2013; Seghier and Price, 2012b). Equally, individual brain networks may be associated with multiple functions; for example, FPCN activation has been implicated during rest and internally-focused states, as well as externally-focused task conditions (Leech et al., 2012a; Spreng, 2012). Therefore, the configuration of functional brain networks may not be static, but context dependent, changing over time when transitioning between cognitive states.

Current methodologies are unable to easily evaluate these complex mechanisms where both the temporal aspects of the interactions between networks and the spatial configuration of those networks can vary with cognitive demands. Specifically, traditional analyses using the regression of psychological time courses onto voxel-wise functional data are not designed to identify coherent functional brain networks from the mixtures that constitute brain wide activation patterns, nor do they capture the relationships that exist between spatial activation patterns across time. In contrast, data-driven approaches, including independent component analysis (ICA) and principal component analysis (PCA), do extract functional networks of brain regions, and can even be used to examine the interactions between networks across task conditions (Allen et al., 2012). However, whilst these data-driven methods capture the complexity of activity over time in terms of multiple networks, they assume that any individual network is spatially static, varying only in activity across time. Consequently, they cannot determine whether the spatial properties of networks evolve in a consistent manner across time and cannot differentiate between these competing perspectives.

Here, we present a novel method for spatiotemporal ICA (stICA) of task FMRI data. We use the technique to examine the spatial evolution of functional brain networks within the context of transitions between cognitive states. To identify components that, for a given temporal window, can vary both in space and time, we propose a simple reorganization of data that can subsequently be used with standard spatial ICA. Specifically, we rearrange adjacent FMRI time-points in space, before performing the ICA, so that the subsequent ICA finds spatiotemporal components (see Figure 1 in Methods). This approach generates components that can evolve spatially over time, thereby providing a simple data-driven method for observing the dynamics of network activity accompanying the onset and offset of a cognitive task, without having to specify the networks *a priori* or assume the spatial distribution of networks is fixed over time. This method allows us to capture two phenomena that can be difficult to capture using a spatial ICA: (i) consistent sequences of transitions between different networks that occur under specific cognitive conditions (e.g., when switching between tasks, performing sequential actions or reacting to a stimuli); (ii) situations in which the spatial configuration of a network changes across time. We explore this approach using both a simple synthetic dataset and empirical FMRI data from a working memory and target detection tasks.

## Methods

### Spatiotemporal independent component analysis

Figure 1 illustrates the data reorganization and analysis involved in the stICA. The approach starts by defining *anchor points*, that are the starting and finishing time-points of periods that are theoretically interesting (Figure 1, top). In the current work, we investigate the onset and offset of task blocks both in simulated and empirical task FMRI datasets (involving interleaved task-rest blocks). Therefore, we define our anchor points as the first time-point and the last time-point of each task block. Given the slow changes in the hemodynamic response, these time-points reliably precede any detectable task-evoked change in the neural signal (i.e., change from rest to task or from task to rest). For each dataset (i.e., each subject or simulated subject), we extracted the whole-brain simulated or empirical FMRI volume at each anchor time-point and the following nine time-points. We then spatially concatenated the 10 volumes extracted for each anchor point (i.e., each task block onset or offset) along the medial-lateral dimension (Figure 1, middle). For the empirical FMRI data this involved taking an image of dimension 23 x 28 x 23 voxels, and creating a sample of dimensions 230 x 28 x 23 voxels. Multiple samples (i.e., 10 volume-wide images) were then temporally concatenated along the fourth dimension (normally referred to as the "temporal dimension" in spatial ICA of FMRI data) (Figure 2, middle). This resulted in an image for the empirical dataset with 230 x 28 x 23 x 1080 voxels (i.e., one 10 x 3D volume x 1080 samples). Finally, the resulting 4D dataset was decomposed into independent components (each size 230 x 28 x 23 voxels) by entering it into probabilistic ICA implemented in Melodic 3.13 (Figure 1, bottom).

The ICA algorithm we use (Melodic 3.13 (Beckmann et al., 2005)) finds spatially independent components. However, since we have reorganized volumes of the same brain acquired at different time-points into the same three dimensions normally used to code for space, the ICA will now find spatiotemporal components: that is, the algorithm is optimizing independence over both space and time. A key assumption is that over the course of the temporal analysis window (i.e. the number of adjacent concatenated volumes), there are patterns that have a temporal dependency (i.e., that extend over multiple adjacent time-points). However, in the situation where there is no consistent temporal structure but consistent spatial structure (e.g., when variation between adjacent volumes is random but there are voxels that co-vary consistently with each other), the resulting components will be restricted to single time-points, and the same spatial pattern of voxels will appear at multiple time-points (see Simulation results below).

We quantitatively assessed the spatial evolution over time of each spatiotemporal component identified. For the empirical FMRI derived components (size 230 x 28 x 23 voxels), we subdivided the image along the medial-lateral dimension to produce 10 volumes (23 x 28 x 23 voxels) corresponding to the within-component time-points. Spatial correlation and mutual information between these volumes was calculated, resulting in a similarity matrix.

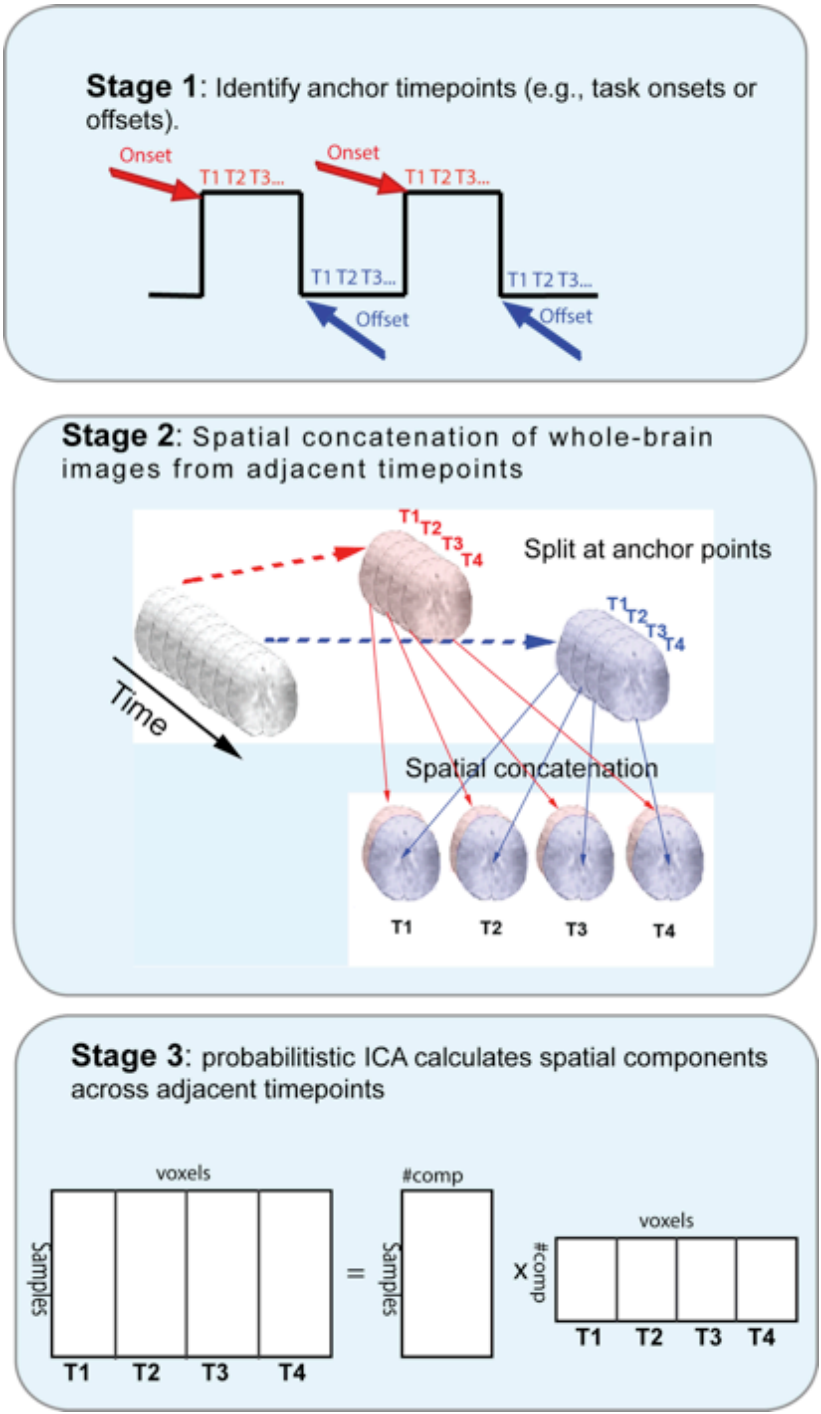

FIGURE 1: Spatiotemporal independent component analysis (stICA). Schematic showing high-level description of the approach.

## Simulated FMRI data

The objective of the simulation was to demonstrate the stICA procedure, how it differs from standard spatial ICA, and show it is able to detect spatiotemporal component when ground truth is known. We performed two simple simulations in MATLAB (R2014a), involving transitions between regions of activation (described below). Both simulations involved the simple representation of brain activity as 3D synthetic data (dimensions 100 x 100 x time). We generated pixel time courses using Gaussian noise (mean = 0, s.d. = 0.2) and activation, deactivation and baseline states were modeled as all voxels within a specified region being "turned on" to 1, -1 or 0, respectively. Regions composed of adjacent 20 x 20 voxels (Figure 3A and 3B, top right) were defined and task timecourses (Figure 2A and 2B, top right) were added to the underlying noise timecourses. The synthetic data were then divided up based on predefined anchor points, spatially concatenated and combined across datasets (as described above). The resulting dataset was then decomposed into spatiotemporal components.

The first simulation (Figure 2A) demonstrates a situation where a given pattern of activation could be followed by two different patterns of activation. It involved three regions and a temporal structure as follows: The activation of Region 1 is followed equally frequently by activity in Region 2 or Region 3, and then both Region 2 and Region 3 are followed by Region 1. This results in four possible transitions: Region 1-> Region 2 and its symmetric transition Region 2-> Region 1; and Region 1-> Region 3 and its symmetric transition Region 3-> Region 1. When a region was fully activated ("on"), its timecourse was set to 1, when fully inactive ("off") it was set to 0. When a region was turned on, its activity linearly increased over 5 TRs to full activation; similarly, when turned off, its activity linearly decreased over 5 TRs to 0. One hundred datasets were generated with 40 TRs per dataset. Given that we implemented slow onset or offsets in the signals (crudely approximating a hemodynamic response function), anchor points were placed at the starts of transitions which were 10 TRs apart (i.e., TR=1, 11, 21, 31), capturing the start states and end states of the transitions. We predicted that the stICA should find two components, with each component reflecting a separate symmetric pair of transitions (one for Region 1-> Region 2 and its reverse; and a second for Region 1-> Region 3 and its reverse).

The second simulation (Figure 2B) explores the situation where there is spatial non-stationarity, and consists of two regions (where one region changes shape consistently with a transition). Region 1 has two spatial patterns, either Region 1+ which contains 20x20 voxels, or Region 1- which contains a subset of 15x15 of the voxels in Region 1+. In the simple simulation, Region 2 is activated, Region 1- is deactivated. In contrast, Region 2 is deactivated and Region 1+ is activated. In this simulation, regions become either fully active (value set to 1) or fully deactivated (value set to -1). For each transition, a region's activity linearly increased or decreased over 5 TRs. One hundred datasets were generated with 30 TRs per dataset, resulting in a 100x100x300 dataset. Anchor points were again set to the start points of the transitions, at TR=1, 11, 21. We predicted that the stICA should find a single component reflecting the changing spatial pattern, with the smaller area corresponding to Region 1- being initially deactivated and Region 2 activated, followed by a spatially larger Region 1+ becoming activated and Region 2 becoming deactivated.

## Empirical FMRI dataset

### Subjects

Functional MRI (FMRI) datasets from 68 healthy control participants (50 female, mean age 30.9 ± 2.76 1sd) alternating between rest (fixation) and either target detection or 2-back working memory task conditions (see below) were obtained from the Human Connectome Project FMRI datasets (Van Essen et al., 2013). Participants provided written informed consent.

### FMRI data acquisition and preprocessing

Four hundred and five gradient-echo echo planar images were collected with whole-brain coverage using an accelerated multiband acquisition protocol (Feinberg et al., 2010; Moeller et al., 2010; Setsompop et al., 2012) with the following parameters: repetition time=720 ms; echo time=33.1 ms; flip angle=52 deg; FOV=208x180 mm (RO x PE); Matrix=104x90 (RO x PE); Slice thickness=2.0 mm; 72 slices with 2.0 mm isotropic voxels; Multiband factor=8; Echo spacing=0.58 ms; and bandwidth=2290 Hz/Px.. For each participant, two runs were acquired, one with phase encoding from right-to-left and the other from left-to-right.

The data was minimally preprocessed according to the standard Human Connectome Project preprocessing pipeline, described in (Fischl, 2012; Glasser et al., 2013; Jenkinson et al., 2002; Jenkinson et al., 2012). In brief, this included: (i) correction for field inhomogeneities, (ii) motion correcting the data; (iii) transforming and resampling the data into 2mm x 2mm x 2mm MNI152 standard space. We down-sampled the data to 8mm x 8mm x 8mm voxel resolution to mitigate the computational requirements of the modified probabilistic ICA step (see below). Additionally, the data were temporally filtered using a 30 second high-pass filter (although qualitatively similar results were obtained without any temporal filtering).

**FMRI stimuli and design**

The tasks are reported in detail by the Human Connectome Project. However, briefly: participants underwent blocks of either target detection (0-back) or 2-back trials interspersed with blocks of rest (fixation). Participants were visually presented with blocks of pictures of places, tools, faces or body parts. Stimuli were projected onto a computer screen behind the participant's head via a mirror approximately 8cm from the participant's eyes. During each run, there were four long blocks separated by approximately 15 seconds of rest (fixation). Each of the long blocks (approximately 50 seconds) consisted of two shorter blocks (approximately 25 seconds each) with either target detection or 2-back stimuli from one of the four categories. At the start of each the short blocks, the task type (2-back or target detection) and target (if target detection) was displayed for 2.5 seconds. This was then followed by 10 trials of 2.5 seconds. Each stimulus was presented for 2 seconds with an inter-trial interval of 500ms. For simplicity, we analyzed only the beginning (placing anchor points at the start of the onset) and end (placing anchor points at the start of the offset) of the long blocks, although future work could explore the transitions between different task states rather than between rest and different tasks as here. Each participant performed two runs of the task. The block presentation order was the same across participants.

**FMRI analysis**

For the empirical FMRI data, we spatially concatenated the 10 volumes extracted for each anchor point (i.e., each task block onset or offset) along the medial-lateral dimension (Figure 2, middle), providing a period (10 x 720 ms = 7.2 s) encompassing the peak of the haemodynamic response function, which occurs about 4–6 s following stimulus delivery (Logothetis, 2002). Here, per subject we have two runs with four task onset samples and four task offset samples for each run. To remove any subject specific mean signal effects, each voxel within the spatially concatenated data was demeaned within a run (in the fourth dimension) before being concatenated across runs and subjects (also in the fourth dimension). This resulted in a single 4D dataset, with 1088 samples (68 subjects x 2 runs x 8 onsets/offsets).

The 4D dataset was then decomposed into different components using spatial probabilistic ICA (Figure 2, bottom). The ICA was decomposed into 20-, 30- and 70- independent components, to investigate if the results were robust to different levels of decomposition, and the consistency of the components calculated at different dimensionality was assessed using spatial correlation. For example, each component derived at 20-dimensions was spatially correlated with each component at 70-dimensions, to assess which, if any, components were preserved in both analyses. As a further validation step, a separate analysis was performed after concatenating 15 rather than 10 adjacent time points, and then analyzing the data using a 30-dimensional ICA. For display purposes, components were thresholded following mixture modeling (as part of the Melodic process) using a probability >0.95 that a voxel was more likely component than noise (the thresholds for each component are presented alongside the supplementary figures that present each component in more detail).

Each component consisted of a spatial map and had an associated set of weights (with one value per 1088 samples). In typical spatial ICA, the weights are considered as the component time course, with a value for each TR, representing how strongly active that component is at that time point. Here, we do not have time points as with a spatial ICA< instead we have a sample for each window (e.g., for each 10 time points), consequently, the stICA results in a weight for each sample, representing how strongly active it is for that sample. These weights can be used to investigate whether a specific component was involved in specific task conditions. To do this, the weight vector (1088 data points for each component) were entered as the dependent variable into a mixed effects general linear model (using fitlme in Matlab 2014). The model contained fixed effect main effects for both task sequence (i.e., onset/offset) and task type (2-back or target detection) and the interaction between the two as well as subject (since there were multiple samples per subject) which was modeled as a random intercept. Overall model F-statistics were calculated to investigate if the component was modulated b y the task (Bonferroni-corrected for multiple comparisons), and t-statistics were calculated for the main effects and for the interactions to see if they were significantly different from the null hypothesis (i.e., no difference).

# Results

## Simulation results

Figure 2A presents the results from the stICA performed on the first simulated dataset. The synthetic data contained four symmetrical transitions (Region 1 <-> Region 2 and Region 1 <-> Region 3). A stICA was calculated with two components. The two components strongly resembled the expected transitions. The first component represented the transition from Region 1 -> Region 2. This is shown as (from left to right) a decline in activation in Region 1 to a negatively activated state, while simultaneously there is an increase in activation in Region 2. This captures the pattern of activity starting at the anchor point at TR 11. Because the underlying transition pattern was symmetric between Region 1 -> Region 2 and Region 2 -> Region 1, this component also captures the reverse transition starting at TR=21 (i.e., if the component is multiplied by -1, it captures the transition from Region 2 -> Region 1. .The second component resembles the transition between Region 1 and Region 3, with again, Region 1 deactivating over time (from left to right) and Region 3 activating. This is consistent with the anchor point starting at TR=31. Again, this component also caputures the reverse transition from Region 3 to Region 1 that starts at TR=1.

We note that the data is demeaned and variance normalized as part of the preprocessing with Melodic; as such the "baseline" inactive state for the different Regions is negative, rather than 0 (as in the inputted timecourse shown in the figure) and the negative "baseline" value is different between Region 1 and Regions 2 and 3, which are active only half as frequently.

We also note that a standard spatial ICA does not capture the transitions in the same form. A spatial ICA on the original data (i.e., performed by running Melodic without the spatial reordering step) revealed three components, corresponding to Regions 1, 2 and 3, but did not directly find the different transitions.

The second simulation (Figure 2B) further demonstrates the approach when there is spatial non-stationarity (i.e., the regional pattern of activation changes with the transition). A stICA was calculated with a single component: theoretically, the minimum necessary to capture the transition. The component displayed spatial non-stationarity: starting (on the left) with a region resembling Region 1- (the reduced size square area) negatively weighted while Region 2 was positively weighted. This then changed over time into Region 1+ (the larger square area) being positively weighted and Region 2 being negatively weighted. We note that a again the standard Melodic spatial ICA failed to produce a component structure that easily captures the ground truth. Figure 2B, bottom, shows the results of a spatial ICA with three components (a number which might be expected to capture all three states: Region 1-, Region 1+ and Region 2). This analysis splits Region 1 into two separate spatial patterns, rather than replicating the spatial organization of Region 1- and Region 1+. The same difficulty of the standard spatial ICA to capture the underlying spatiotemporal structure (because of the non-stationarity of Region 1) occurs with 1, 2, 5, or 10 component analyses.

## A: Multiple transitions

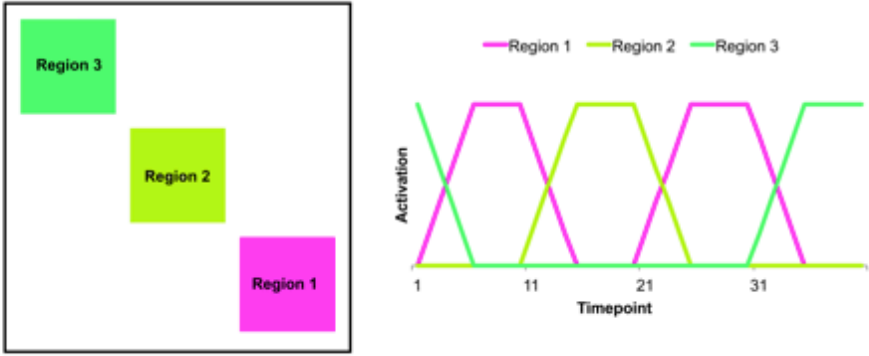

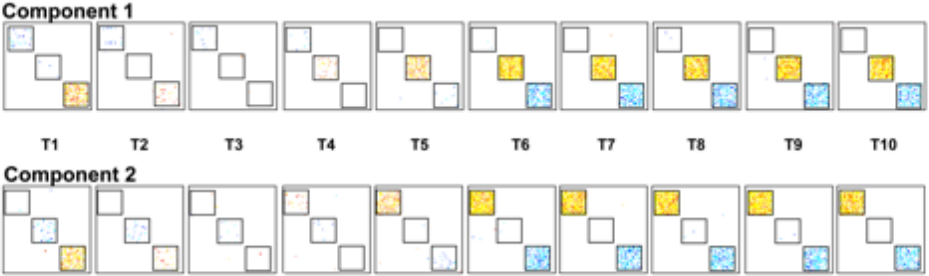

## B: Spatial non-stationarity

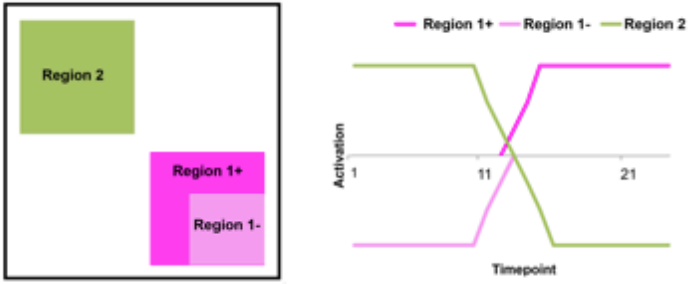

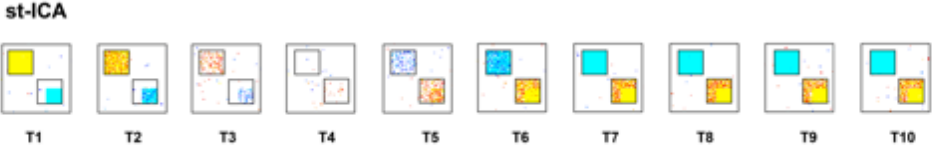

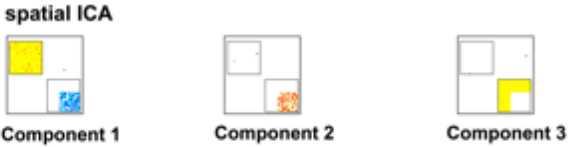

*FIGURE 2: Simulated FMRI data, method and results of spatiotemporal independent component analysis (stICA).* Two examples of simple synthetic data illustrating two different examples of transitions that the stICA approach can directly detect. A: two symmetric transitions where Region 1's activity precedes either Region 2 or 3 with equal frequency, or Regions 2 or 3 precede Region 1. Top left is an illustration of the spatial patterns used to generate the synthetic data. Top right is an illustration of the timecourses used, with the spatial patterns, to create transitions. Below are the components found from the stICA. B: A second simulation, illustrating a case of spatial non-stationarity. Region 1 and Region 2 are anti-correlated; however, Region 1 is a different size when it is deactivated (Region 1-) than when it is activated (Region 1+). At the bottom of 2B we see, first, the stICA that captures the spatial non-stationarity, and an example of a standard spatial ICA that, in this case, fails to capture the spatial non-stationarity. For both simulations, voxels are displayed at z-values where the probability of a voxel being part of the component rather than noise was >0.95. To aid visualization, square black boxes have been placed around the distinct regions in the ICA output.

## Empirical results

We performed the stICA on the empirical FMRI data, extracting 30 components. Nine components are presented in Figure 3 (more detailed presentations of the components are included in the Supplementary Materials). A mixed effects general linear model comparing each component's loadings with the task condition (i.e., whether a sample was working memory or target detection and whether it was an onset or offset) showed that eight components significantly affected by the task (Bonferroni correcting for multiple comparisons). Two components were positively weighted with task onsets and six with task offsets. In addition, we present another component (Component 7) that, although not being modulated by the task (e.g., onset/offset) is reported here because it has a strong resemblance to the canonical default mode network (DMN). The other 21 components are presented in Supplementary Figure 9.

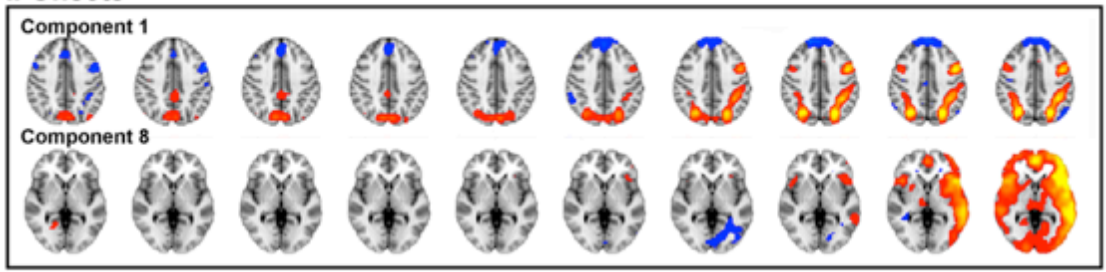
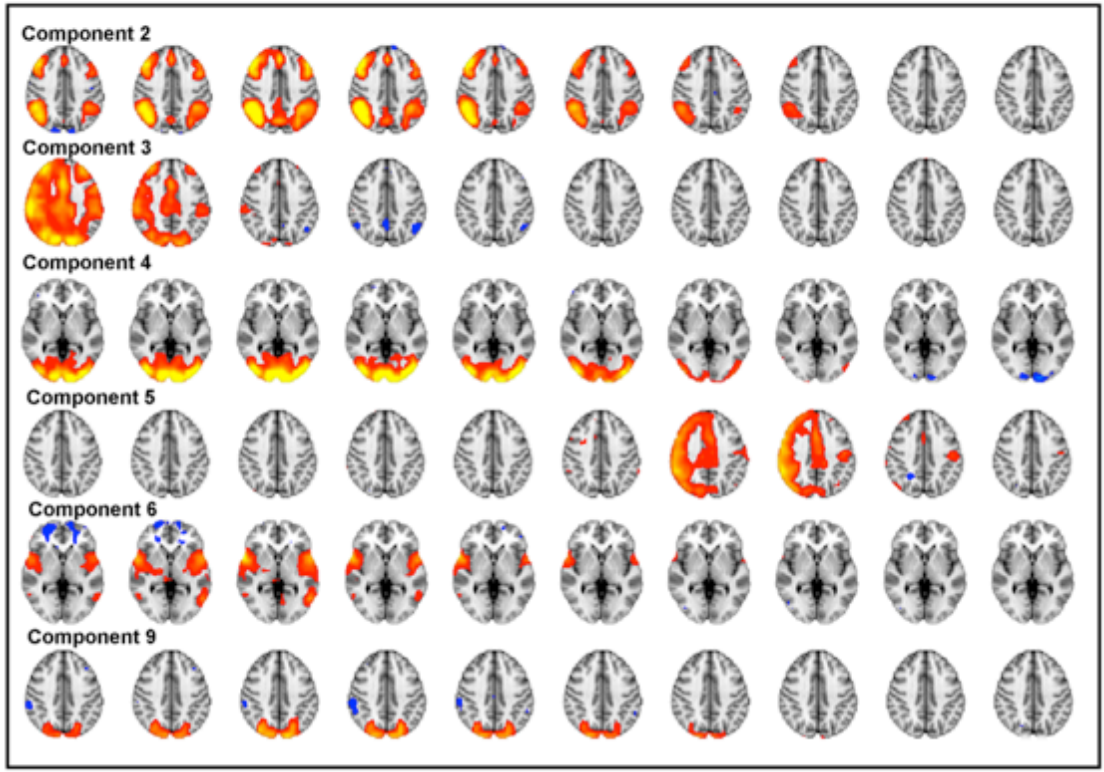
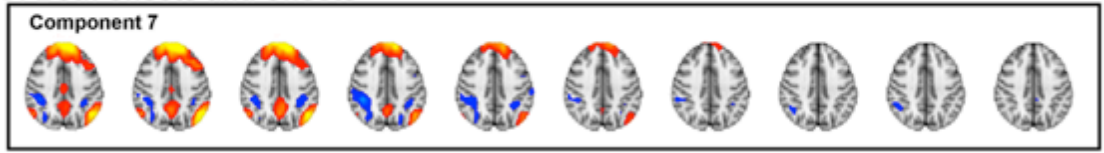

**FIGURE 3: Nine spatiotemporal components extracted from a 30-component spatiotemporal independent component analysis (stICA).** *Time is presented from left to right, different columns representing the 10 time-points following an anchor point. Cool colors represent voxels negatively coupled to the component and warm colors positively coupled. All components were classified according to whether they were significantly*

*(Bonferroni-corrected for multiple comparisons) more activated for task onset anchor points, (A) or offsets (B), or neither (C).*

Within Component 1, we see over time (i.e. from left to right in Figure 3) the component changes dramatically, consistent with a shift from being in a resting state to a state performing an externally focused cognitive task. The component weights revealed that it was strongly weighted to the onset anchor points, i.e. a transition from rest to task. Figure 4 shows the temporal evolution in the spatial pattern within each component (by using spatial correlation and mutual information between time-points within the component). We see that Component 1 evolves such that the first and last time-points are highly spatially anti-correlated (although, the anti-correlation is not perfect and, as we see below in Figure 5B, the spatial patterns are non-stationary, changing with the transition).

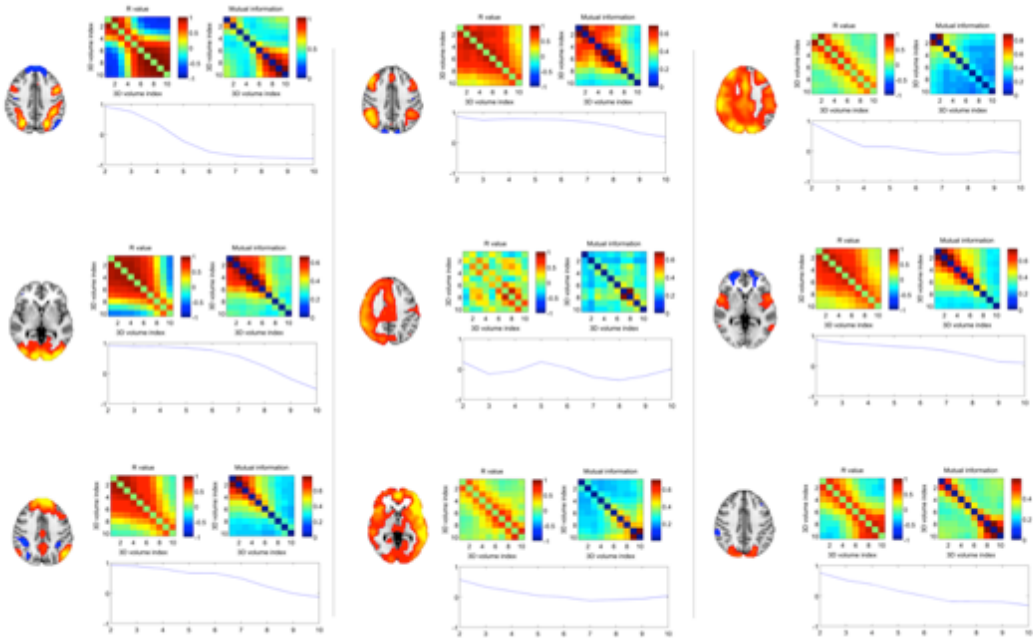

**FIGURE 4: Assessing spatial change over time within spatiotemporal components.** *For each component the similarity matrix (calculated using spatial correlation and mutual*

*information) between each time-point is presented. In addition, the spatial correlation between the first time-point and each additional time-point is also presented. Brain images shown are a representative time-point from within each spatiotemporal component. Full components are shown in Figure 3 and Supplementary Material.*

The majority of the offset components showed the reverse temporal evolution from Component 1. Components 2, 4, 6 and 9 displayed a shift over time from typically task-positive networks of brain regions being active to being inactive (Figure 3B, Components 2, 4, 6 and 9). As with Component 1, this progression is consistent with the temporal change predicted based on a canonical hemodynamic response model of a task offset. These findings suggest the modified ICA has decomposed the offset periods into multiple different well-recognized functional networks. Component 2 is composed of bilateral superior parietal, lateral and medial superior frontal regions, a variant on fronto-parietal control networks. Components 4 and 9 are visual networks localized to predominantly occipital regions. As with Component 1, Figure 4 shows these components shifting over time to a spatially anti-correlated state. Component 6 is composed of bilateral anterior insula/frontal operculum and medial frontal regions, consistent with another typical network involved in cognitive control (Bonnelle et al., 2012; Menon and Uddin, 2010).

Of the remaining components, Component 7 strongly resembles the canonical DMN and anti-correlated parts of the dorsal attention network, displaying a gradual reduction in activity over time. Surprisingly, this network was not significantly more active in task onsets (where a reduction of DMN might be expected to accompany task engagement) than offsets ($p=0.37$, uncorrected for multiple comparisons).

Three components, 3, 5 and 8, partially resembled right or left-lateralized fronto-parietal control networks (although extending over much of the grey matter in the cortex). Component 5 is only transiently present (for three time-points) which is inconsistent with the expected extended hemodynamic temporal smoothing of the neural signal, and appears in the middle of the time window, it seems likely that this component may be non-neural in origin (Figure 3). Further, the spatial correlation and mutual information of the network is not consistent with a smooth transition between states (Figure 4). As such, Component 5 may be non-neural in origin. Components 3 and 8 are also more transient than predicted based on the typical hemodynamic response (Figure 3, 4) as such it is unclear whether they are neural or non-neural, or possibly a mixture of the two.

The spatiotemporal evolution of Component 1 is presented in more detail in Figure 5. At time-point one (T1, Figure 5A) there are positively coupled medial frontal and parietal regions (reminiscent of the medial nodes of the DMN), and negatively coupled occipital, superior parietal, pre-supplementary motor area and dorsolateral pre-frontal regions (regions typically activated by externally-focused cognitive tasks). This pattern evolves over time, with the expected fronto-parietal system becoming apparent from T5 (approximately 3.5 s from the task onset) and being fully established by T7 (approximately 5 s post onset). This is consistent with the timing of the peak of the canonical hemodynamic response function (about 4-6 s) (Logothetis, 2002).

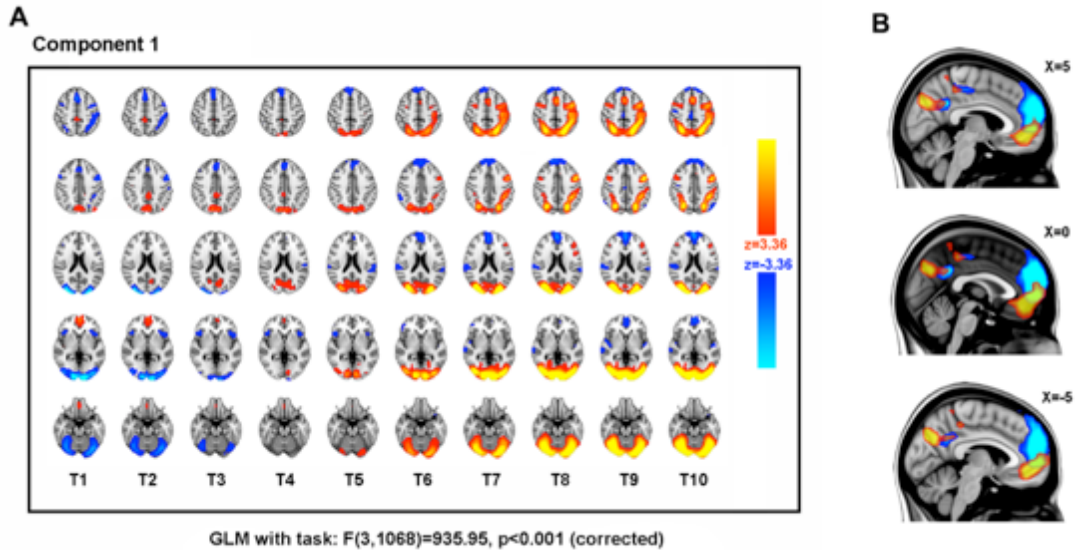

**FIGURE 5: The first non-noise spatio-temporal component identified by a 30-component spatiotemporal independent component analysis (stICA).** *(Component 1 in Figure 3). The component was strongly weighted to onset anchor points, i.e. a transition from rest to task. (**A**) The component is shown in multiple axial planes (top to bottom, MNI Z co-ordinate = 50, 40, 20, 0 and -20), with time presented from left to right (T1 to T10). Cool colors represent voxels negatively coupled to the component and warm colors positively coupled. (**B**) The evolving spatial organization of the medial default mode network regions of the component is shown for time-points T1 and T10, with voxels coupled to the component at the first time-point T1 (thresholded z>2) shown in red, and voxels within the final time-point (T10) in blue (component thresholded z<-2). The three panels show different X-coordinates in MNI space.*

In general, Component 1 captures the change in spatial pattern expected for a transition between the resting state and an active externally-focused cognitive task. However, the component also suggests that, rather than the network resembling the DMN simply changing from a relatively 'active' state to a relatively 'less active' state, its spatial organization also changes (Figure 5B). In particular, we see that in the first time-point (Figure 5B, red), the DMN extends posteriorly from the posterior cingulate and precuneus and shows activation in areas of ventral anterior cingulate and para-cingulate frontal regions. In contrast, by time-point 10 (Figure 5B, blue), the network is now relatively deactivated and we see that, while there are areas of overlap with the first-time-point, the parietal regions extend anteriorly, and the frontal regions extend into superior frontal regions. The same evolution within Component 1 was seen at a range of thresholds for component inclusion, and when comparing other time-points, e.g., T2 with T9. Supplementary Figure 10 shows the difference between T1 and T10, illustrating that the non-stationarity we observe is not merely a function of the threshold chosen.

Up to now, we have only considered how the components are weighted differently for task onset and offset. However, task blocks were composed of two different externally-focused cognitive tasks: either a target detection (i.e. 0-back) or a 2-back task. Both tasks had a similar composition, with a visually presented picture stimulus cueing a forced choice cognitive decision followed by a motor response (button press). Both tasks had the same inter-trial and inter-stimulus timings. However, the tasks differed in terms of their cognitive demands, with target detection requiring a single target picture to be maintained in memory across the whole block, while the 2-back required the continuous reorganised and updating of items in working memory. Consistent with their different cognitive demands, the two tasks showed a difference in terms of behavior, with the target detection being performed more accurately and more rapidly (mean of median accuracy = 0.92, s.d.= 0.09; mean of median reaction time = 807ms, s.d., 134) than the 2-back task (mean of median accuracy = 0.82, s.d.= 0.12; mean of median reaction time = 1028ms , s.d. = 137).

Figure 6 shows how different components are loaded for the two tasks, and for the onsets and offsets of the two tasks. Table 1 presents results from 2x2 mixed effects general linear models (with a random intercept accounting for subject) performed for each component, with task type (target detection or 2-back) and task onset (block onset or offset) as main effects and also including a task type by onset interaction. We found that task onset/offset was the most significant predictor, but that for most components there were significant differences between task-type. Of the components weighted heavily for task onset, Component 1 (the network that transitions from a default-mode type network to a fronto-parietal control network) was significantly more heavily involved with the onset of the target detection than the onset of the 2-back task. For the components involved in the offset of tasks, Component 2 (a fronto-parietal network) was more involved with the offset of the 2-back than the target detection task, as was Component 3.  Component 8 had a significant overall model F statistic (see Supplementary Figure 8) but neither of the main effects nor the interaction was significant on its own.

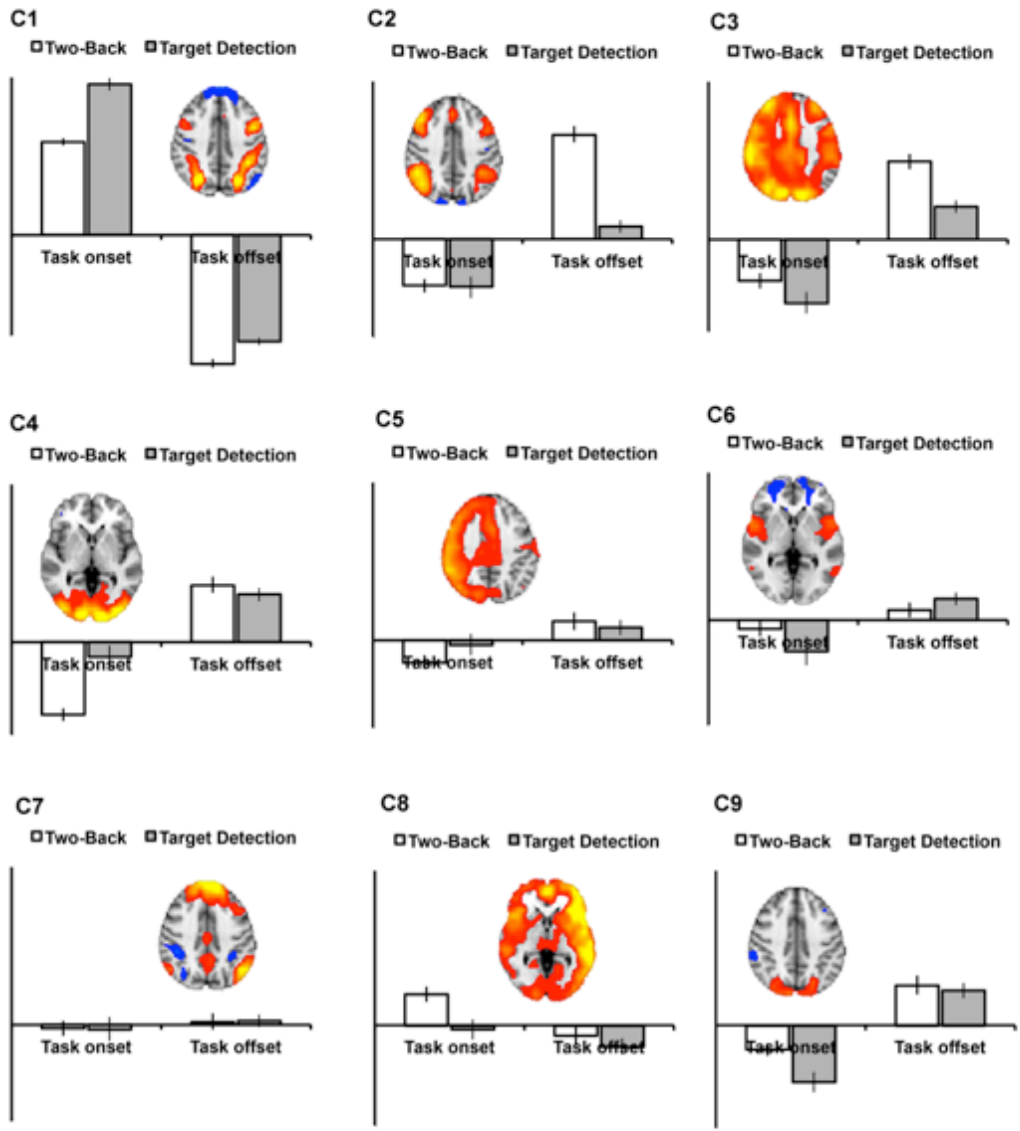

FIGURE 6: **Different components' involvement in onset and offset of different types of cognitive task.** Each of the components are displayed alongside their mean weighting for onsets/offsets for the 2-back and target detection tasks. The y-axis is in arbitrary units. Error bars are standard errors of the mean. Brain images shown are a representative time-point from within each spatiotemporal component.

Table 1:

| Component | Condition | T | p (one tailed) |
|---|---|---|---|
| C1 | 2-back vs Target Detection | -3.45 | <0.005 |
|  | Onset vs Offset | 39.94 | <0.0001 |
|  | Interaction | -3.85 | <0.005 |
|  |  |  |  |
| C2 | 2-back vs Target Detection | 8.27 | <0.0001 |
|  | Onset vs Offset | -5.44 | <0.0001 |
|  | Interaction | -5.77 | <0.0001 |
|  |  |  |  |
| C3 | 2-back vs Target Detection | 3.97 | <0.005 |
|  | Onset vs Offset | -8.48 | <0.0001 |
|  | Interaction | -1.45 | ns |
| C4 | 2-back vs Target Detection | 0.87 | ns |
|  | Onset vs Offset | -5.56 | <0.0001 |
|  | Interaction | -4.24 | <0.001 |
| C5 | 2-back vs Target Detection | 0.49 | ns |
|  | Onset vs Offset | -1.44 | ns |
|  | Interaction | -1.35 | ns |
| C6 | 2-back vs Target Detection | -0.87 | ns |
|  | Onset vs Offset | -4.04 | <0.0005 |
|  | Interaction | 1.83 | ns |
| C7 | 2-back vs Target Detection | -0.075 | ns |
|  | Onset vs Offset | -0.68 | ns |
|  | Interaction | 0.10 | ns |
| C8 | 2-back vs Target Detection | 0.90 | ns |
|  | Onset vs Offset | 1.40 | ns |
|  | Interaction | 1.36 | ns |
| C9 | 2-back vs Target Detection | 0.46 | ns |
|  | Onset vs Offset | -7.57 | <0.0001 |
|  | Interaction | 1.59 | ns |

*Results from 2x2 mixed effects general linear models performed for each component, with task type (target detection or 2-back) and task onset (block onset or offset) as main effects and also including a task type by onset interaction.*

Given the probabilistic nature of the ICA and its sensitivity to the number of components extracted, we re-ran the original analysis, extracting 20 and 70 components. The comparison of the 20- and 70-dimensional analyses with the original 30-component analysis are presented in Figure 7A. We found a very close correspondence between Component 1 from the original analysis (Figure 7A, middle row) with the two other analyses (Figure 7A, top and bottom). When using spatial correlation to quantitatively compare across the components extracted, there were very strong correspondences for the majority of the nine components of interest (Figure 7B). Finally, to show that the results were tolerant to different numbers of adjacent time-points being concatenated, Figure 7C and D (in comparison with equivalent components shown in Figure 3) show the very strong similarity in specific components between analyses conducted with 15 time-points and the original analysis with 10 time-points. We note that within Component 1, the medial default mode network structures (i.e., posterior cingulate and ventromedial prefrontal cortex, cold colors) that are present in the original 10-time-point analysis do not survive the conservative statistical threshold used in the 15-time-point analysis. However, at lower thresholds these regions are present.

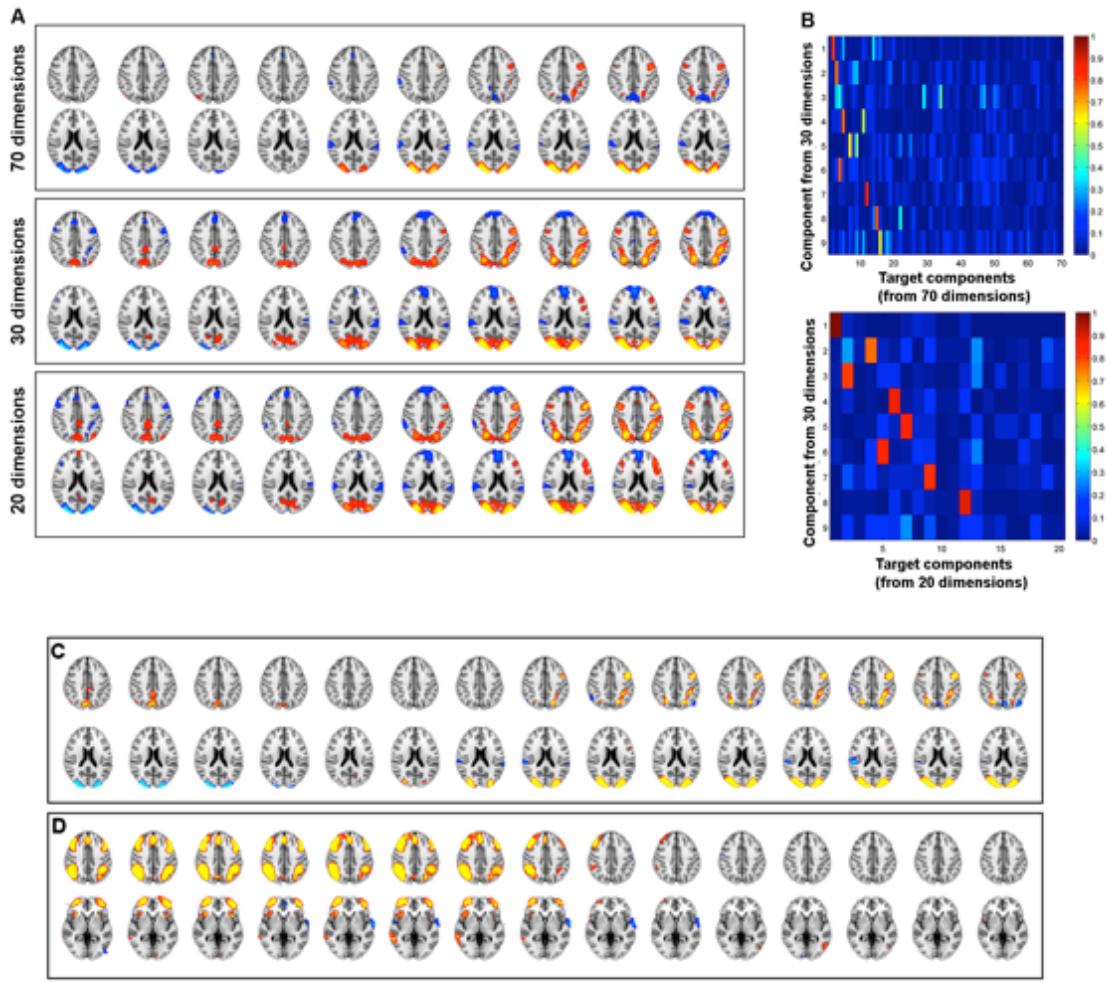

**FIGURE 7: Demonstrating the robustness of the findings with different analyses.**
*(A)* The homologue of Component 1 from the 30-component stICA (middle row) in both a 20- (bottom) and a 70-component (top) analysis. *(B)* Matrix of spatial correlation coefficients between each of the nine non-noise components from the 30-component stICA and components from either the 20- (top matrix) or 70-dimension analyses (bottom). This shows strong, unique correlations between most of the nine components and homologues in the 20- and 70-dimension analyses. *(C and D)* Exploring the results when spatially concatenating 15 rather than 10 time-points, using a 30-dimensional stICA. Time is from left to right. **(C)** The homologue of Component 1 (see Figure 3A for comparison) **(D)** The homologue of Component 2 (see Figure 3B).

# Discussion

Here, we found our simple adaptation to a frequently used data-driven approach robustly defines transitions both in synthetic data and, from empirical FMRI data, that involves well-recognized functional networks that accompany the onset and offset of a cognitive task. The approach explicitly focuses on spatiotemporal transitions rather than temporal or spatial patterns, and allow for both spatial and temporal non-stationarity and for spatial patterns to be involved in multiple different transitions. From a cognitive perspective, our approach captures transitions between brain networks, which can allow us to investigate the mechanisms underlying cognitive transitions that are fundamental to efficient and flexible behavior. Our results are consistent with current understanding of the broad dynamical pattern based on previous studies: a default mode network (DMN) that becomes less active while a set of fronto-parietal control networks (FPCN) become more active with the onset of task. However, in our first exploration of this novel approach, we have observed a number of more subtle changes with potentially important consequences for understanding network dynamics.

Although this work is a primarily a proof of concept study, there are several interesting neurobiological findings that the analysis reveals in addition to the expected transition between DMN and FPCN with task. First, the DMN has sometimes been described as a spatially consistent and stable network; however, increasingly it is seen to be more reactive to task demands (e.g., (Harrison et al., 2008) (Leech et al., 2011; Leech et al., 2014; Seghier and Price, 2012a)). However, here, we see the DMN that is anti-correlated with the FPCN has a dynamically changing spatial distribution with cognitive state. The DMN shifts to a more anterior/superior pattern of relative deactivation during task from the more posterior inferior pattern of activation during rest. More work is needed to understand the functional significance of this spatial variation, but one possibility is that the DMN and other functional networks are spatially dependent and may be constantly shifting their spatial organization as some form of ongoing macroscopic homeostatic plasticity (Leech et al., 2014). In a similar vein, a second DMN component was identified, consistent with the DMN playing multiple functional roles (Leech et al., 2012b; Leech and Sharp, 2013). Further, this DMN component, which showed activated DMN regions reducing in activation over time, was equally weighted to task onsets and offsets; that is, there was a 'reducing' DMN component following the onset of task (as typically expected) but also following the end of task and into rest. This suggests that the transition to rest from task is not necessarily a completely passive process. Instead, it may be more of an active process, consistent with recent results suggesting a transient increase in DMN activity following task (Cocchi et al., 2013).

Our analysis also suggests an asymmetry between task onset and offset: i.e., transitioning to rest is not the same as transitioning into task. There were two components that accompanied the onset of the task: (i) a DMN-type component that transitioned into fronto-parietal and visual networks (that typically activate during these types of task (Bonnelle et al., 2011; Sharp et al., 2011)); (ii) an extensive network covering much of the grey matter but centered on frontal executive systems including bilateral insula and anterior cingulate regions (which, given its late onset and extensive activation pattern, may be noise). However, the offset of task was followed by multiple components showing a 'reducing' pattern (i.e., activation reducing over time). Components C4 and C9, which both show initially positive coupling of the visual regions to the component, were significantly associated with task offsets and not task onset. Within the components there was reducing activation over time and eventual deactivation (negative component coupling) in C4. This is as expected for the end of a visual task moving into rest. One possible explanation for the existence of relatively few onset components compared to multiple different offset components is that transitioning from an unconstrained rest state to a cognitive task involves an initial transition into a highly constrained neural state that is relatively consistent across both tasks. Over time during the task block, more heterogeneous neural processes may become evident, possibly because the relatively homogenous set of neural processes necessary to establish a new cognitive set are no longer necessary, allowing a more unconstrained or metastable state (Hellyer et al., 2014a) or because of variability in how task performance is sustained over time (Leech et al., 2014).

The transitions evoked by the onsets or offsets of the two different tasks (i.e. target detection or 2-back) also differed. Although components were generally weighted to being overall task onset or offset, there was also a smaller influence from the type of task, such that Component 1, although involved for both tasks, was more heavily weighted for the target detection task. Similarly, for offsets, Component 2 (a second fronto-parietal control network) was more heavily weighted for the 2-back. These findings may reflect shared neural systems supporting similarities in the cognitive operations involved in both tasks. But they also imply that the 2-back task may involve additional cognitive resources, shown as the engagement during onset (and the disengagement during offset) of additional fronto-parietal systems. Differences between onsets and offsets and between tasks may also reflect heterogeneity across subjects in how they performed the task. This is something that can be studied in the future by looking at datasets where there is considerable individual variability in task strategy or performance e.g., in patient groups or more executive tasks that elicit a variety of behavioral strategies.

Our stICA approach modifies a well-accepted spatial ICA procedure by rearranging adjacent time-points in space, before performing the ICA on samples concatenated across subjects. This approach generates components that are not constrained in space or time (except by the size of the rearranging time window), but rather span both dimensions. By anchoring the analyses to data from time periods associated with the onset and offset of a cognitive challenging task, our approach reveals novel aspects of the dynamics of network activity accompanying the transition into and out of task conditions. Importantly, these findings were observed without having to specify *a priori* either the spatial networks or the specific task time courses (specifying instead only the anchor points) and allowing the spatial distribution of networks to evolve over time.

Whilst our simulations were highly simplified, they enabled us to demonstrate some important aspects of our stICA approach, supporting our application of the approach to empirical FMRI data. The simulations also allowed comparison with standard spatial ICA where ground truth was known. We found stICA was able to directly capture two phenomena defined within the synthetic data: (i) a consistent transition between networks and (ii) non-stationarity of networks over time. We found that standard spatial ICA (implemented in Melodic) in both cases failed to produce components that captured the underlying spatiotemporal structure of the data as clearly. To uncover this structure with spatial ICA, additional analysis steps using the component timecourse would be necessary, and accurately pulling apart the appropriate underlying transition structure would depend on the choice of additional analysis approach.

Our approach differs in important ways from existing approaches to investigating functional brain networks during cognitive tasks. In the traditional mass univariate voxel-wise approach, a time-course known to be associated with the task is regressed with functional data, to find voxels which are active during task and rest. This produces a single map of regions that are associated with the time-course, but may miss the possibility that regions associated with the task are the result of activity or competition between or within multiple functional networks over time. More recently, data-driven approaches, such as spatial ICA, have been used to define networks of brain regions, and the time-course of networks have been compared with a time-course associated with the task. Comparing the output of our stICA procedure to standard spatial ICA is non-trivial. For instance, the model order of a spatial ICA would need to be far higher to capture the same rich description of spatiotemporal transitions. Increasing the model order comes with problems of multiple comparisons and in interpretation, and this approach still assumes stationarity of spatial pattern of components, which may not be the case.

While we have focused on using ICA to find spatiotemporal components within FMRI data, a similar rearranging approach (rearranging time into space) could also be combined with other functional connectivity approaches, and could be used with other functional data. For example, alternative clustering approaches such as PCA could be used. Indeed, seed-based approaches could also be used by selecting a voxel from the first time-point, and maps computed based on voxels that significantly covary with this voxel at the same time-point and across different time shifts could be calculated. However, such mass univariate approaches would fail to take full advantage of the multivariate nature of the data, and the fact that there are likely to be multiple different temporally evolving components depending on task, within-individual or across-individual differences.

The approach we have taken here is not the only way to find spatiotemporal components. Previously suggested ICA strategies (e.g., convolutive ICA (Cocchi et al., 2013)) can create spatiotemporal components, and comparing ICAs and other data clustering approaches generated from time-windowed data uncovers low frequency spatial dynamics in brain networks (Akemann et al., 2010; Braga et al., 2013; Hellyer et al., 2014b). Similarly, there are other approaches such as the temporal functional modes approach that combines spatial and temporal ICA (Smith et al., 2012), although the temporal functional modes approach does not allow the spatial maps to evolve over time. However, the approach we present is highly tractable (involving simple spatial rearranging of the data), and as such has considerable potential as an exploratory data analysis tool. In principle, the approach can be applied to any FMRI data, for example, in resting state data where the anchor points can not be determined by knowledge of an external task, anchor points for rearranging can be determined using for example, k-means clustering. By using anchor points for the rearranging we can constrain the ICA to find components that are focused specifically on parts of the dataset that are biologically interesting, whereas performing the analysis without these anchors, while possible, would be less constrained and so less likely to find coherent structure in the data.

# References

Akemann, W., Mutoh, H., Perron, A., Rossier, J., Knopfel, T., 2010. Imaging brain electric signals with genetically targeted voltage-sensitive fluorescent proteins. Nat Methods 7, 643-649.

Allen, E.A., Damaraju, E., Plis, S.M., Erhardt, E.B., Eichele, T., Calhoun, V.D., 2012. Tracking whole-brain connectivity dynamics in the resting state. Cerebral Cortex.

Beckmann, C.F., DeLuca, M., Devlin, J.T., Smith, S.M., 2005. Investigations into resting-state connectivity using independent component analysis. Philos Trans R Soc Lond B Biol Sci 360, 1001-1013.

Bonnelle, V., Ham, T.E., Leech, R., Kinnunen, K.M., Mehta, M.A., Greenwood, R.J., Sharp, D.J., 2012. Salience network integrity predicts default mode network function after traumatic brain injury. Proc Natl Acad Sci U S A 109, 4690-4695.

Bonnelle, V., Leech, R., Kinnunen, K.M., Ham, T.E., Beckmann, C.F., De Boissezon, X., Greenwood, R.J., Sharp, D.J., 2011. Default Mode Network Connectivity Predicts Sustained Attention Deficits after Traumatic Brain Injury. The Journal of neuroscience : the official journal of the Society for Neuroscience 31, 13442-13451.

Braga, R.M., Sharp, D.J., Leeson, C., Wise, R.J., Leech, R., 2013. Echoes of the Brain within Default Mode, Association, and Heteromodal Cortices. The Journal of Neuroscience 33, 14031-14039.

Buckner, R.L., Andrews-Hanna, J.R., Schacter, D.L., 2008. The brain's default network: anatomy, function, and relevance to disease. Ann N Y Acad Sci 1124, 1-38.

Chang, C., Glover, G.H., 2010. Time–frequency dynamics of resting-state brain connectivity measured with fMRI. Neuroimage 50, 81-98.

Cocchi, L., Zalesky, A., Fornito, A., Mattingley, J.B., 2013. Dynamic cooperation and competition between brain systems during cognitive control. Trends in cognitive sciences 17, 493-501.

Dosenbach, N.U., Fair, D.A., Miezin, F.M., Cohen, A.L., Wenger, K.K., Dosenbach, R.A., Fox, M.D., Snyder, A.Z., Vincent, J.L., Raichle, M.E., Schlaggar, B.L., Petersen, S.E., 2007. Distinct brain networks for adaptive and stable task control in humans. Proceedings of the National Academy of Sciences of the United States of America 104, 11073-11078.

Erhardt, E.B., Allen, E.A., Wei, Y., Eichele, T., Calhoun, V.D., 2012. SimTB, a simulation toolbox for fMRI data under a model of spatiotemporal separability. Neuroimage 59, 4160-4167.

Feinberg, D.A., Moeller, S., Smith, S.M., Auerbach, E., Ramanna, S., Glasser, M.F., Miller, K.L., Ugurbil, K., Yacoub, E., 2010. Multiplexed Echo Planar Imaging for Sub-Second Whole Brain FMRI and Fast Diffusion Imaging. PLoS ONE 5, e15710.

Fischl, B., 2012. FreeSurfer. Neuroimage 62, 774-781.

Fox, M.D., Snyder, A.Z., Vincent, J.L., Corbetta, M., Van Essen, D.C., Raichle, M.E., 2005. The human brain is intrinsically organized into dynamic, anticorrelated functional networks. Proceedings of the National Academy of Sciences of the United States of America 102, 9673-9678.

Glasser, M.F., Sotiropoulos, S.N., Wilson, J.A., Coalson, T.S., Fischl, B., Andersson, J.L., Xu, J., Jbabdi, S., Webster, M., Polimeni, J.R., Van Essen, D.C., Jenkinson, M., 2013. The minimal preprocessing pipelines for the Human Connectome Project. Neuroimage 80, 105-124.

Harrison, B.J., Pujol, J., Contreras-Rodríguez, O., Soriano-Mas, C., López-Solà, M., Deus, J., Ortiz, H., Blanco-Hinojo, L., Alonso, P., Hernández-Ribas, R., Cardoner, N.,

Menchón, J.M., 2011. Task-Induced Deactivation from Rest Extends beyond the Default Mode Brain Network. PLoS ONE 6, e22964.

Harrison, B.J., Pujol, J., López-Solà, M., Hernández-Ribas, R., Deus, J., Ortiz, H., Soriano-Mas, C., Yücel, M., Pantelis, C., Cardoner, N., 2008. Consistency and functional specialization in the default mode brain network. Proceedings of the National Academy of Sciences 105, 9781-9786.

Hellyer, P.J., Shanahan, M., Scott, G., Wise, R.J., Sharp, D.J., Leech, R., 2014a. The Control of Global Brain Dynamics: Opposing Actions of Frontoparietal Control and Default Mode Networks on Attention. The Journal of Neuroscience 34, 451-461.

Hellyer, P.J., Shanahan, M., Scott, G., Wise, R.J., Sharp, D.J., Leech, R., 2014b. The control of global brain dynamics: opposing actions of frontoparietal control and default mode networks on attention. The Journal of neuroscience : the official journal of the Society for Neuroscience 34, 451-461.

Jenkinson, M., Bannister, P., Brady, M., Smith, S., 2002. Improved optimization for the robust and accurate linear registration and motion correction of brain images. Neuroimage 17, 825-841.

Jenkinson, M., Beckmann, C.F., Behrens, T.E., Woolrich, M.W., Smith, S.M., 2012. FSL. Neuroimage 62, 782-790.

Leech, R., Braga, R., Sharp, D.J., 2012a. Echoes of the brain within the posterior cingulate cortex. J Neurosci 32, 215-222.

Leech, R., Braga, R., Sharp, D.J., 2012b. Echoes of the brain within the posterior cingulate cortex. The Journal of neuroscience : the official journal of the Society for Neuroscience 32, 215-222.

Leech, R., Kamourieh, S., Beckmann, C.F., Sharp, D.J., 2011. Fractionating the default mode network: distinct contributions of the ventral and dorsal posterior cingulate cortex to cognitive control. The Journal of neuroscience : the official journal of the Society for Neuroscience 31, 3217-3224.

Leech, R., Scott, G., Carhart-Harris, R., Turkheimer, F., Taylor-Robinson, S.D., Sharp, D.J., 2014. Spatial Dependencies between Large-Scale Brain Networks. PLoS ONE 9, e98500.

Leech, R., Sharp, D.J., 2013. The role of the posterior cingulate cortex in cognition and disease. Brain : a journal of neurology.

Logothetis, N.K., 2002. The neural basis of the blood-oxygen-level-dependent functional magnetic resonance imaging signal. Philos Trans R Soc Lond B Biol Sci 357, 1003-1037.

Menon, V., Uddin, L.Q., 2010. Saliency, switching, attention and control: a network model of insula function. Brain structure & function 214, 655-667.

Moeller, S., Yacoub, E., Olman, C.A., Auerbach, E., Strupp, J., Harel, N., Ugurbil, K., 2010. Multiband multislice GE-EPI at 7 tesla, with 16-fold acceleration using partial parallel imaging with application to high spatial and temporal whole-brain fMRI. Magn Reson Med 63, 1144-1153.

Raichle, M.E., MacLeod, A.M., Snyder, A.Z., Powers, W.J., Gusnard, D.A., Shulman, G.L., 2001. A default mode of brain function. Proceedings of the National Academy of Sciences of the United States of America 98, 676-682.

Seghier, M.L., Price, C.J., 2012a. Functional heterogeneity within the default network during semantic processing and speech production. Front Psychol 3.

Seghier, M.L., Price, C.J., 2012b. Functional Heterogeneity within the Default Network during Semantic Processing and Speech Production. Front Psychol 3, 281.

Setsompop, K., Gagoski, B.A., Polimeni, J.R., Witzel, T., Wedeen, V.J., Wald, L.L., 2012. Blipped-controlled aliasing in parallel imaging for simultaneous multislice echo planar imaging with reduced g-factor penalty. Magnetic Resonance in Medicine 67, 1210-1224.

Sharp, D.J., Beckmann, C.F., Greenwood, R., Kinnunen, K.M., Bonnelle, V., De Boissezon, X., Powell, J.H., Counsell, S.J., Patel, M.C., Leech, R., 2011. Default mode network functional and structural connectivity after traumatic brain injury. Brain : a journal of neurology 134, 2233-2247.

Shmuel, A., Yacoub, E., Pfeuffer, J., Van de Moortele, P.-F., Adriany, G., Hu, X., Ugurbil, K., 2002. Sustained negative BOLD, blood flow and oxygen consumption response and its coupling to the positive response in the human brain. Neuron 36, 1195-1210.

Smith, S.M., Fox, P.T., Miller, K.L., Glahn, D.C., Fox, P.M., Mackay, C.E., Filippini, N., Watkins, K.E., Toro, R., Laird, A.R., Beckmann, C.F., 2009. Correspondence of the brain's functional architecture during activation and rest. Proceedings of the National Academy of Sciences of the United States of America 106, 13040-13045.

Smith, S.M., Miller, K.L., Moeller, S., Xu, J., Auerbach, E.J., Woolrich, M.W., Beckmann, C.F., Jenkinson, M., Andersson, J., Glasser, M.F., Van Essen, D.C., Feinberg, D.A., Yacoub, E.S., Ugurbil, K., 2012. Temporally-independent functional modes of spontaneous brain activity. Proceedings of the National Academy of Sciences of the United States of America 109, 3131-3136.

Spreng, R.N., 2012. The fallacy of a "task-negative" network. Front Psychology, 145.

Van Essen, D.C., Smith, S.M., Barch, D.M., Behrens, T.E.J., Yacoub, E., Ugurbil, K., 2013. The WU-Minn Human Connectome Project: An overview. Neuroimage 80, 62-79.

Vincent, J.L., Kahn, I., Snyder, A.Z., Raichle, M.E., Buckner, R.L., 2008. Evidence for a frontoparietal control system revealed by intrinsic functional connectivity. J Neurophysiol 100, 3328-3342.